# Evidence for an asymptotic giant branch star in the progenitor system of a type Ia supernova

Mario Hamuy*¶, M. M. Phillips†, Nicholas B. Suntzeff‖, José Maza‡, L. E. González‡, Miguel Roth†, Kevin Krisciunas†, Nidia Morrell†, E. M. Green§, S. E. Persson*, & P. J. McCarthy*

*Carnegie Observatories, 813 Santa Barbara St., Pasadena, CA, USA*

*† Las Campanas Observatory, Carnegie Observatories, Casilla 601, La Serena, Chile*

*‖ Cerro Tololo Inter-American Observatory, National Optical Astronomy Observatories, Casilla 603, La Serena, Chile*

*‡ Departamento de Astronomía, Universidad de Chile, Casilla 36-D, Santiago, Chile*

*§ University of Arizona, Steward Observatory, Tucson, AZ 85721, USA*

*¶ Hubble Fellow*

**A type Ia supernova (SN Ia), one of the two main classes of exploding stars, is recognized by the absence of hydrogen and the presence of elements such as silicon and sulphur in its spectra. These explosions are thought to produce the majority of iron-peak elements in the universe and are known to be precise "standard candles" used to measure distances to galaxies. While there is general agreement that SNe Ia are exploding white dwarfs[1], astronomers face the embarrassing problem that the progenitor systems have never been directly observed. Significant effort has been put into the detection of circum-stellar material (CSM) in order to discriminate between the different types of possible progenitor systems[2], yet no CSM has been found[3]. Here we report optical observations of SN 2002ic which reveal large amounts of CSM seen as a strong hydrogen emission. This observation**

**suggests that the progenitor system contained a massive asymptotic branch giant star which lost a few solar masses of hydrogen-rich gas prior to the type Ia explosion.**

SN 2002ic was discovered before or near maximum light by the Nearby Supernova Factory search[4]. It appeared ~4 arc seconds west of a faint elongated galaxy and north of a fainter galaxy (see Supplementary Information figure 1). We have measured a redshift of z=0.22 for the galaxy to the west, which rules out any association with the supernova (z=0.07; see below). No spectrum is yet available for the fainter galaxy to the south of the supernova.

Our spectroscopic coverage of SN 2002ic encompasses a period of ~60 days, and clearly demonstrates that this object belongs to the Ia class (see Fig. 1). The features in SN 2002ic are very much like those of the type Ia SN 1991T[5], but diluted in strength. Although SN 1991T/1999aa-like events are sometimes referred to as "peculiar", they make up 20% of the local population of SNe Ia[6]. The most remarkable spectroscopic feature in SN 2002ic is the Hα emission at z=0.0666. This emission is spatially unresolved with a FWHM ≤1.2 arcsec (1.7 kpc for $H_0$=65 km s$^{-1}$ Mpc$^{-1}$). The Hα profiles (Fig. 2) have an unresolved (FWHM < 300 km s$^{-1}$) component on top of a broad (FWHM ~ 1800 km s$^{-1}$) base. While the narrow component could be an unrelated H II region, the broader component cannot be explained in this way. Such narrow+broad Hα profiles have no precedent among SNe Ia[3], but are typical of SNe IIn[7] where it is thought that the narrow lines arise from the un-shocked CSM photo-ionized by radiation from the SN, while the intermediate width lines are formed in the shock-heated CSM[8,9]. The origin of the CSM in SNe IIn is attributed to the SN progenitor, a presumably massive star that undergoes mass loss before explosion. By analogy, the double component Hα profile in SN 2002ic is clear evidence for a dense CSM. The measured Hα fluxes in SN 2002ic correspond to energies of ~2x10$^{40}$ erg s$^{-1}$ in each of the broad and narrow line components. Photo-ionization models for the narrow component[3] imply an unexpectedly high mass-loss rate $\sim 10^{-2.4}$ $M_\odot$ yr$^{-1}$. The absence of a rapid dimming in Hα or a change in line width (Fig. 2) suggests that the SN/CSM interaction remains strong at least 2 months after explosion, which is also consistent with large mass-loss rates[3].

The SN/CSM interaction in SNe IIn produces a luminosity enhancement compared to classical SNe II[10] due to the conversion of kinetic energy into continuum radiation. This suggests that the absorption features and light curves of SN 2002ic may be similarly "veiled" relative to "normal" SNe Ia. This hypothesis is confirmed from the spectroscopic comparison between SN 2002ic and the type Ia SN 1999ee[11] (Fig. 3), and the photometric analysis shown in Fig. 4.

The strength of the SN/CSM interaction in SN 2002ic is totally unexpected for a SN Ia, but is typical in SNe IIn. It is interesting to hypothesize that some SNe IIn may actually be SNe Ia like SN 2002ic, but with an even stronger SN/CSM interaction. In this context, we draw attention to the type IIn SN 1997cy that may have been associated with the γ-ray burst source GRB 970514[12,13]. In Fig. 5 we show that the late-time (~70 day) spectra of SN 2002ic and SN 1997cy are strikingly similar. Both the BV light curve decline rates (~0.7mag/100d) and the peak absolute magnitude of M(V)~-20.1 of SN1997cy[13] are very similar to the decline rates inferred for the SN/CSM interaction and the peak brightness for SN 2002ic (see Fig. 4). Note that similar high luminosities and flat, constant-colour light curves were also seen in the SN IIn 1988Z[14].

The photometric properties of SN 1997cy are well reproduced by a model of the explosion of a 25 $M_\odot$ star with high explosion energy ($3 \times 10^{52}$ erg) which interacts with a dense CSM of ~5 $M_\odot$[13]. In this model, the light curve is powered by the SN/CSM interaction with a contribution of radioactive heating of up to 0.7 $M_\odot$ of $^{56}$Ni. In light of the close similarity between SN 1997cy and SN 2002ic, and the clear evidence that SN 2002ic was a bona fide SN Ia, a re-examination of possible models for SN 1997cy seems worthwhile including both its association to GRB 970514 and the overall energetics. We note that C/O Chandrasekhar white dwarfs are very unlikely to produce more than $2 \times 10^{51}$ erg[15]. This upper limit appears to be in conflict with the total radiated energy of $10^{52}$ erg estimated for the SN IIn 1988Z[16], but this estimate is highly uncertain.

Is there evidence for a SN/CSM interaction in other SNe Ia? It has been observed that the BVRI light curves of the 1991T/1999a type[6] show an "over-brightness" ~1 month after maximum compared to spectroscopically "normal" SNe Ia[17,18]. Perhaps all 1991T/1999aa events occur in progenitors with a significant CSM, and the SN/CSM interaction accounts for the additional luminosity at later epochs. A test of this hypothesis would be to search for radio or x-ray emission from these SNe Ia or extremely faint hydrogen emission.

The modelling of SN 1997cy shows that a few solar masses of CSM material are involved in the emission processes. By analogy, we would expect a star in the progenitor system of SN 2002ic to lose a similar amount of mass, ruling out a double-degenerate model for the progenitor of this type Ia explosion. A progenitor consistent with this amount of mass loss is a binary system containing a C/O white dwarf and a massive (3-7 $M_\odot$) asymptotic giant branch (AGB) star where the integrated mass loss can reach a few solar masses[19]. The accretion of part of this wind onto the white dwarf could bring it to the Chandrasekhar mass. An alternate explanation is the thermonuclear explosion of the degenerate core of a "single" AGB star in a "type 1.5" event[20], although the spectrum of this event has not been calculated in detail. We also note that SN 2002ic was much more luminous than its (yet-to-be-identified) host galaxy. Curiously, the host galaxy of SN 1997cy was also of very low luminosity[12], as was the host of the 1999aa-like SN 1999aw[18]. Such low-luminosity galaxies are likely to be characterized by low metallicities[21]. Hence, the presence of SN/CSM interaction in 1991T/1999aa-like events such as SN 2002ic may be related to mass loss in a metal-poor environment.

**Supplementary Information** accompanies the paper on **www.nature.com/nature**.

**Acknowledgements** All of the co-authors participated in gathering the observations of the supernova. M.H. noticed the presence of hydrogen emission. M.M.P. noticed spectroscopic and photometric peculiarities and put forward the idea that these could be understood as due to SN/CSM interaction. N.B.S. provided the arguments about the progenitor types. M.H., M.M.P., and N.B.S. co-wrote the paper.

**Correspondence** and requests should be addressed to M.H. (e-mail: mhamuy@ociw.edu).

**Figure 1** Spectroscopic evolution of SN 2002ic. **a**, This sequence shows five spectra of SN 2002ic (in AB magnitudes) obtained between 2002 Nov. 29 and 2003 Feb. 1 UT with the Las Campanas Observatory Baade 6.5-m and du Pont 2.5-m telescopes, and the Steward Observatory Bok 2.3-m telescope. Arbitrary offsets have been added to the spectra for clarity. The spectra are +6, +10, +34, +47, and +70 days from estimated maximum light. We attempted to remove the two most prominent telluric lines (indicated with the circled plus signs symbols), but some residuals are evident. The top spectrum shows the Si II λ6355 feature that defines the Ia class, as well as prominent Fe III absorption features at 4200 and 4900 Å. The absence of the He/Na feature at 5900 Å in the spectral evolution rules out a type Ib/c classification. **b**, A comparison between the 29 Nov 2002 (+6 days) observation of SN 2002ic and the spectrum of the type Ia SN 1991T[5] obtained at an epoch of +4 days, shows that both spectra are quite similar, except that the features in SN 2002ic are all diluted in strength.

**Figure 2** Decomposition and evolution of the Hα profiles. The observed Hα profiles are shown in red for four different epochs. These profiles cannot be fit with single Gaussian models, but two Gaussians of different widths do provide acceptable fits. The black dotted lines show the individual components (with FWHM ~300 and ~1800 km s$^{-1}$) and the black solid lines the sum of both Gaussians. This figure suggests that the Hα profile does not evolve rapidly with time. It is difficult to make more precise statements because, while the subtraction of the continuum at early epochs is straightforward, at later times the SN has a broad emission feature (see Fig. 1a) which makes the subtraction much more uncertain.

**Figure 3** Spectral analysis of SN 2002ic. **a,** Spectrum of SN 2002ic taken on 2002 Dec. 3 UT (heavy black line) compared with that of the normal type Ia SN 1999ee[11] taken 6 days past maximum (thin black line). In red is shown the sum (by eye) of a low-order continuum (blue line) and the spectrum of SN 1999ee. The most prominent telluric lines are indicated with the circled plus signs symbols. **b,** Same as above, except comparing the spectrum of SN 2002ic obtained on 2002 Dec. 27 UT with that of SN 1999ee taken 34 days past maximum. **c,** Same as above, except comparing the spectrum of SN 2002ic

taken on 2003 Jan. 9 UT with that of SN 1990N obtained 47 days past maximum. Except for the features of Ca II at 3950 Å and 8600 Å which appear weaker in SN 2002ic, in all three cases the continuum + SN spectra and the SN 2002ic spectra match extremely well. This simple model explains why the absorption features in SN 2002ic appear "veiled" relative to normal SNe Ia.

**Figure 4** Photometric analysis of SN 2002ic. With filled circles are shown the BVI light curves of SN 2002ic obtained with the Las Campanas 2.5-m and 1-m telescopes. These light curves were derived from differential measurements relative to several local standards calibrated on two photometric nights. Also shown with the "+" symbols in **b** are the unfiltered discovery magnitudes[4]. Maximum light occurred around Nov 23 (JD 2452602) and the peak magnitudes were B~17.7, V~17.4, and I~17.2. Correcting for a Galactic reddening of E(B-V)=0.073[22] and K corrections[23] we derive absolute magnitudes of M(B)~-20.1, M(V)~-20.3, and M(I)~-20.4 ($H_0$=65), which are ~1 magnitude brighter than the most luminous SNe Ia[24]. For comparison are plotted the B and V light curves of the "normal" type Ia SN 1999ee[25] [characterized by a post-maximum decline rate, Δm15(B), of 0.94] expected for the redshift and reddening of SN 2002ic. While the initial decline rate of SN 2002ic was very slow, the B and V light curves evolved into the final linear decline rate within ~25 days of maximum, which is only observed for the *fastest* declining SNe Ia[26]. Also shown as diamonds in B and V is the luminosity evolution of the SN/CSM interaction as inferred from the dilution of the spectral features (Fig. 3) to which we have fit straight lines (shown as dashes). The triangles show the result of subtracting this light curve from the observed photometry of SN 2002ic. For the first ~25 days, the "unveiled" peak magnitudes of B~V~18.0 are consistent with a "normal" SN Ia. However, beyond JD 2452630 the residual flux remains high compared to SN 1999ee. This suggests that a model of the light curves of SN

2002ic as the sum of a "normal" SN Ia and a slow-declining SN/CSM interaction may be overly simple.

**Figure 5** Spectroscopic comparison between SN 2002ic and SN 1997cy. Spectrum of SN 2002ic taken on Jan. 9 (~47 days after maximum light, which corresponds to ~67 days after explosion for an assumed time of 20 days between explosion and peak brightness) compared to that of the type IIn SN 1997cy taken 71 days after explosion[13], which is assumed to coincide with the detection of GRB 970514[12]. The striking similarity between these two objects suggests that some SNe IIn are the result of thermonuclear explosions of white dwarfs surrounded by a dense CSM instead of core collapse in massive stars.

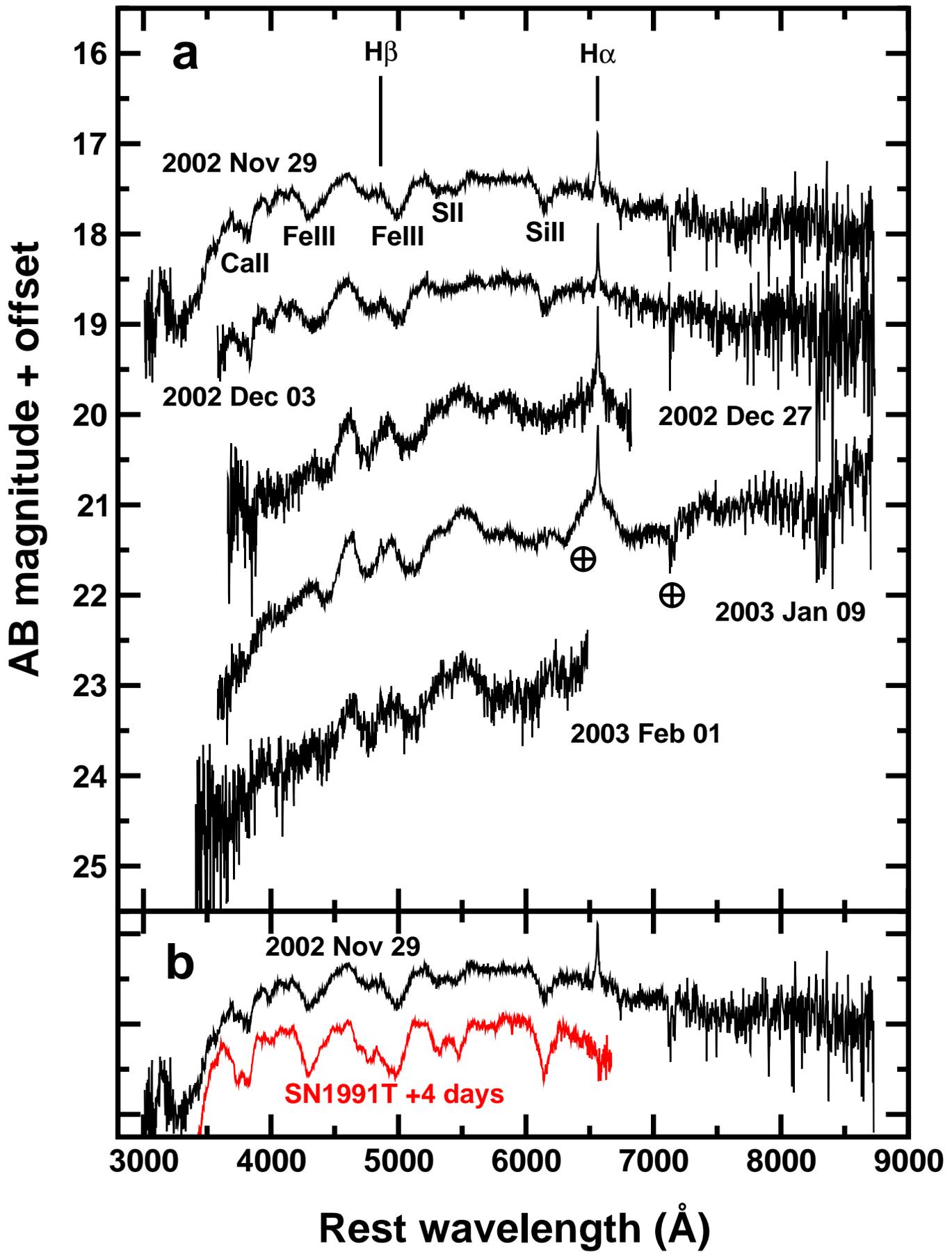

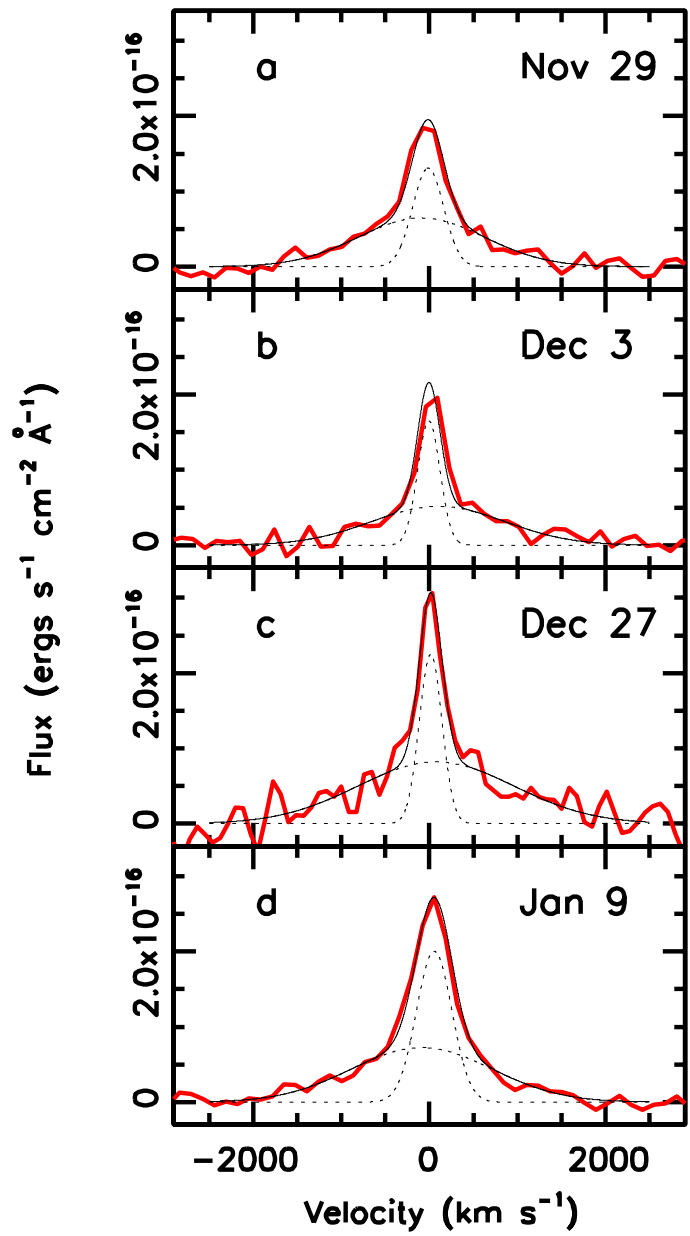

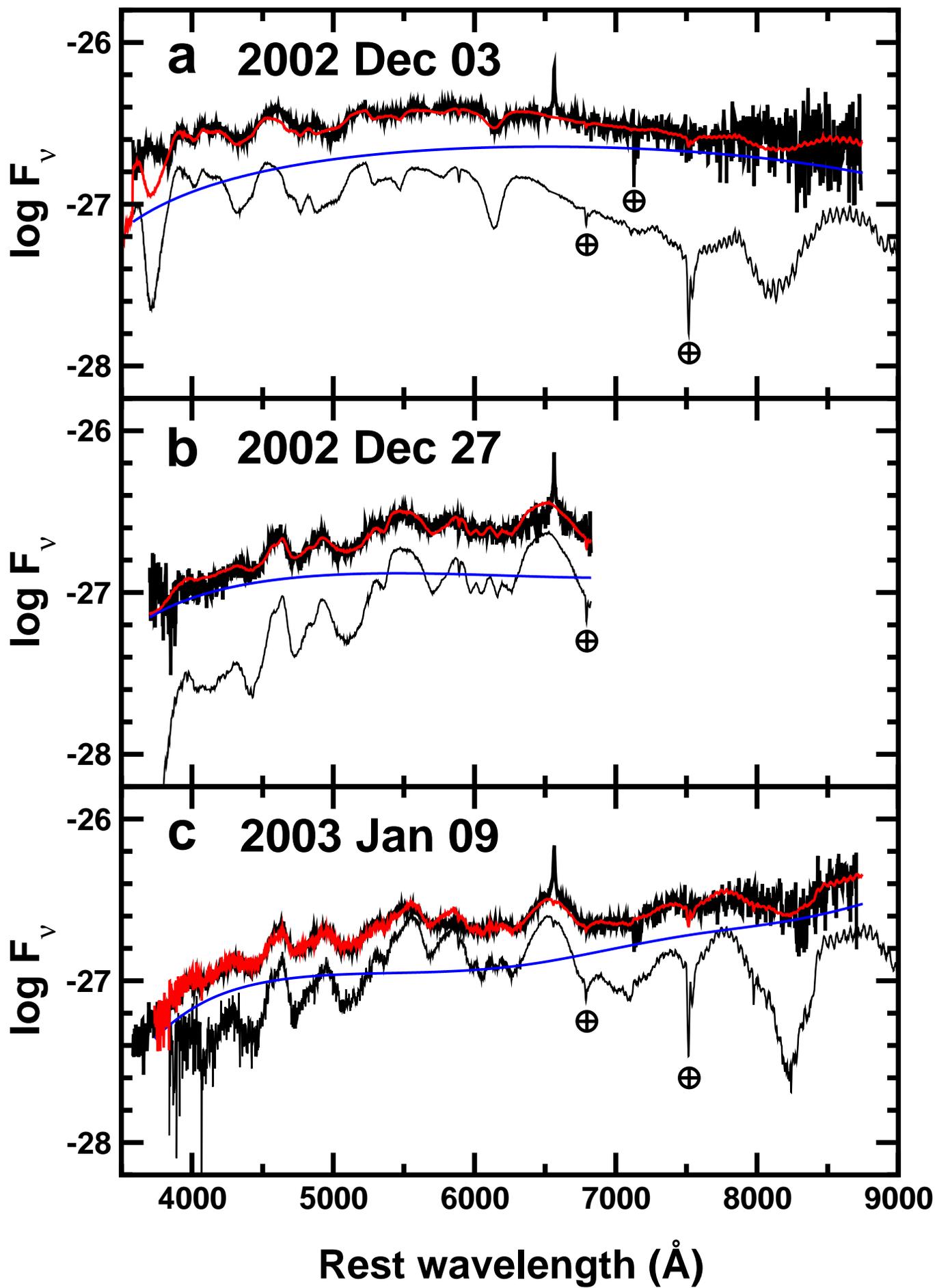

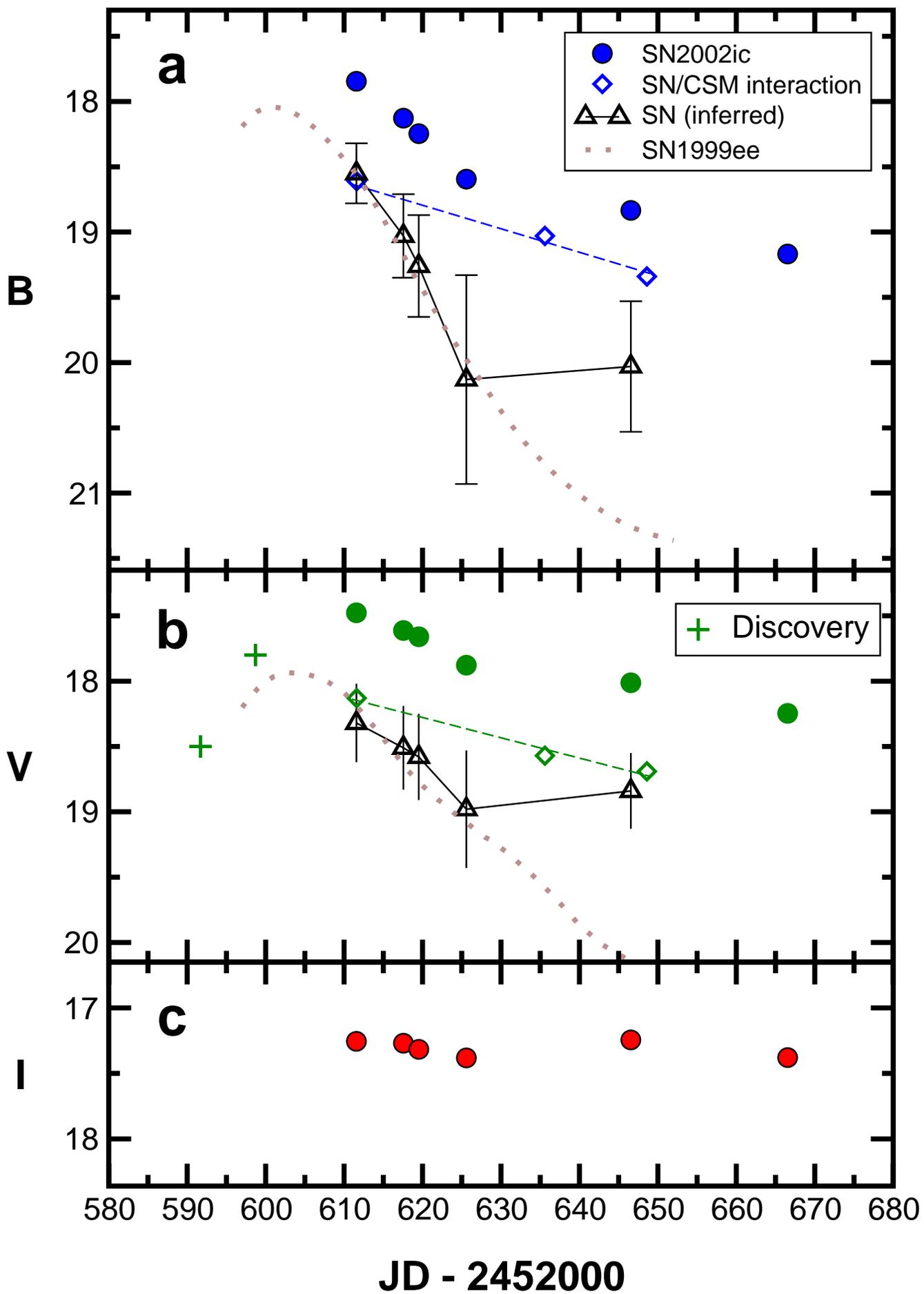

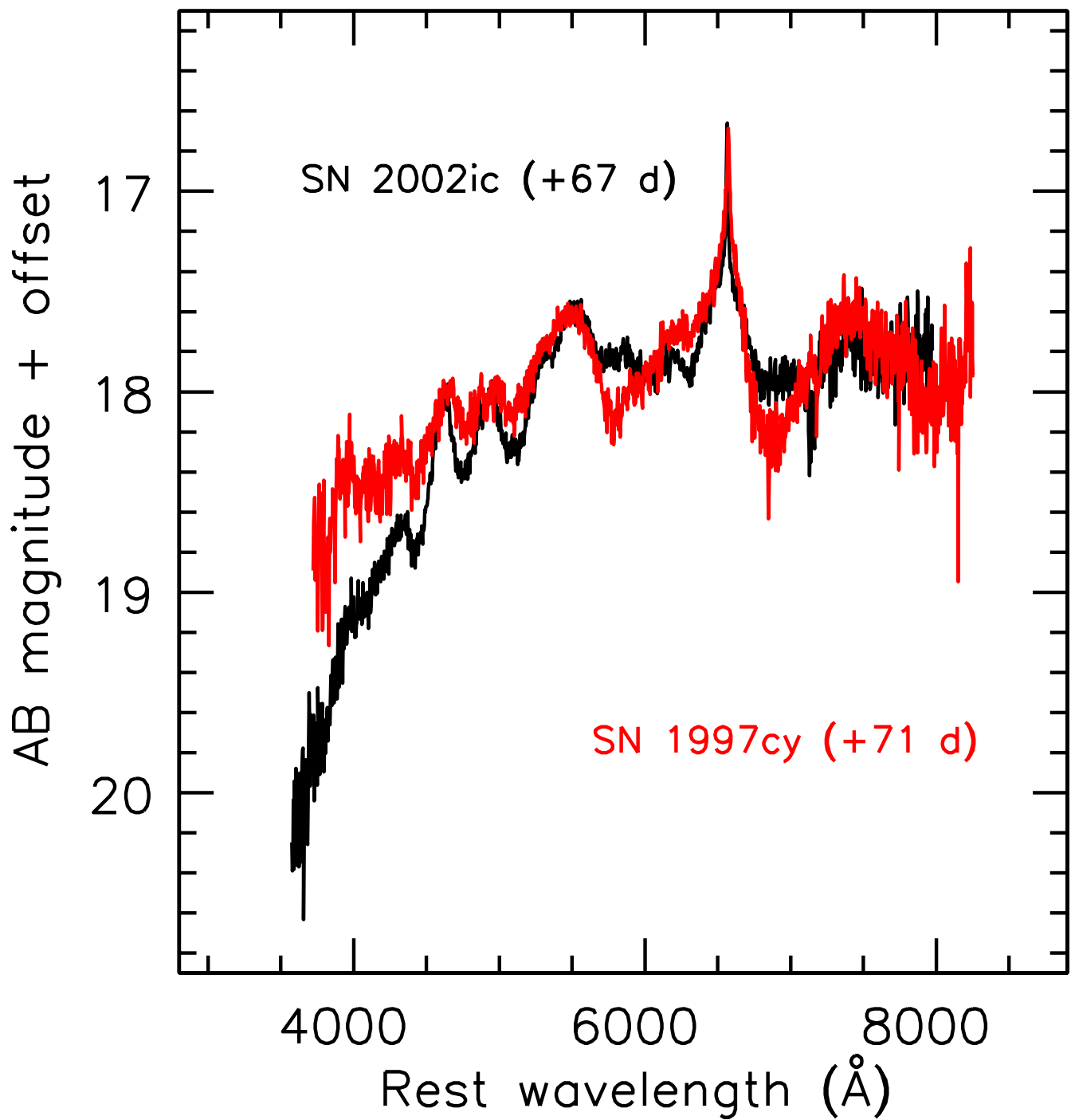